\documentclass[12pt,amssymb,amsmath,tightenlines]{article}
\usepackage{amsmath, amssymb, amsthm, latexsym, mathrsfs,exscale}
\usepackage{dsfont}
\usepackage{graphicx,epic,cite}
\usepackage[usenames,dvipsnames]{color}
\newcommand{\half}{\mbox{$\textstyle \frac{1}{2}$}}

\newcommand{\re}{\mbox{$\rm e$}}

\newcommand{\rd}{\mbox{$\rm d$}}
\numberwithin{equation}{section}

\title{\bf{Modelling Information Flows in Financial Markets}}
\begin{document}

\author{Dorje C. Brody${}^{1}$, 
Lane P. Hughston${}^{1}$ \&  
Andrea Macrina${}^{2,3}$
}
\date{}
\maketitle
\begin{center}
${}^{1}$ Department of Mathematics, Imperial College London, 
London SW7 2AZ \\
${}^{2}$ Department of Mathematics, King's College London, 
London WC2R 2LS  \\
${}^{3}$ Institute of Economic Research, Kyoto University, 
Kyoto 606-8501, Japan
\end{center}

\begin{abstract}
This paper presents an overview of information-based asset 
pricing. In the information-based approach, an asset is defined 
by its cash-flow structure. The market is assumed to have access 
to ``partial" information about future cash flows. Each cash flow is 
determined by a collection of independent market factors called 
$X$-factors. The market filtration 
is generated by a set of information processes, each of 
which carries information about one of the $X$-factors, and 
eventually reveals the $X$-factor in a way that ensures 
that the associated cash flows have the correct measureability 
properties. In the models considered each information process 
has two terms, one of which contains a ``signal" about the 
associated $X$-factor, and the other of which represents ``market 
noise". The existence of an established pricing kernel, adapted to 
the market filtration, is assumed. The price of 
an asset is given by the expectation of the discounted cash flows in 
the associated risk-neutral measure, conditional on the information 
provided by the market. When the market 
noise is modelled by a Brownian bridge one is able to construct 
explicit formulae for asset prices, as well 
as semi-analytic expressions for the prices and greeks of 
options and derivatives. In particular, option price 
data can be used to determine the information flow-rate parameters 
implicit in the definitions of the information processes. One 
consequence of the modelling framework is a specific 
scheme of stochastic volatility and correlation processes. Instead 
of imposing a volatility and correlation model upon the dynamics of a 
set of assets, one is able to deduce the dynamics of the volatilities and 
correlations of the asset price movements from more primitive assumptions 
involving the associated cash flows. The 
paper concludes with an examination of situations involving 
asymmetric information. We present a simple model for informed 
traders and show how this can be used as a basis for so-called 
statistical arbitrage. Finally, we consider the problem of price 
formation in a heterogeneous market with multiple agents. 
\end{abstract}

\section{Cash flow structures and market factors}

In financial markets, the revelation of information is the 
most important factor in the determination of the price movements of  
financial assets. When a new piece of information (whether true, 
partly true, misleading, or bogus) circulates in 
a financial market, the prices of related assets move in response, 
and they move again when the information is updated. But how do 
we build specific models that incorporate the impact of information 
on asset prices? In this article we present an overview of some of 
the key issues involved in modelling the flow of information in 
financial markets, and develop in some detail some elementary 
models for ``information" in various situations. 
We show how information flow processes, when appropriately 
modelled, can be used to determine the associated price 
processes of financial assets. Applications 
to the pricing of various types of contingent claims will also be 
indicated. One of the contributions of the present work is to 
introduce a model for dynamic correlation in the situation where 
we consider a portfolio of assets. Rather than imposing an artificial 
correlation structure on the assets under consideration, we are able 
to infer the correlation structure from more basic assumptions. 
In the final section of the paper, we make some remarks 
about statistical 
arbitrage strategies, and about price formation in markets 
characterised by inhomogeneous information flows. 

When models are constructed for the pricing
and risk management of complicated financial products, the price
dynamics of the simpler financial assets, upon which the more 
complicated products are based, are often simply  ``assumed" (modulo
some parametric or functional freedom). One can understand from 
a practical angle why it can be expeditious to proceed on that basis. 
Nevertheless, from a fundamental view we have to consider that 
even the basic
financial assets (shares, bonds, etc) are characterised by a number 
of potentially ``complex" features, and so to make sense of the 
behaviour of such assets we need to consider 
what goes into the 
determination of their prices. To build up models for 
the dynamics of asset prices, it seems logical to proceed 
step by step along the following lines: 
(1) model the cash-flows arising from the asset as 
random variables; (2) model the market filtration (the flow of 
information to the market); (3) model the pricing kernel (which 
takes into account discounting, risk aversion, and the absence of 
arbitrage); and (4) work out the resulting dynamics for the 
price process.

We model the unfolding of chance in a financial market with the 
specification of a probability space $(\Omega,{\mathcal F}, \mathbb{P})$ 
on which we are going to construct a filtration 
$\{\mathcal{F}_t\}$ representing the flow of information to market 
participants. Here $\mathbb{P}$ denotes the ``physical" 
probability measure. The markets we consider will, in general, be 
incomplete. That is to say, although derivatives can be priced we 
do not assume that they can be hedged. Since we are going to 
model the filtration we say that we are working in an 
information-based asset pricing framework. The general approach 
that we describe here is that of Brody, Hughston \& Macrina 
\cite{[1],[2],[3]}.

Consider a financial instrument that delivers to its owner a 
set of random 
cash flows $\{D_{T_k}\}_{k=1,...,n}$ on the dates $\{T_k\}_{k=1,...,n}$. 
For simplicity, we assume that these dates are fixed, and finite 
in number. The extension to random dates and to an infinite number 
of dates is straightforward. 
Let the pricing kernel be denoted $\{\pi_t\}$. 
At time $t$ the value $S_t$ of a contract that generates the cash 
flows $\{D_{T_k}\}_{k=1,\dots,n}$ is given by the following 
\textit{valuation formula}:
\begin{eqnarray}\label{eq:4.1}
S_t=\frac{1}{\pi_t}\sum^n_{k=1}{\mathds 1}_{\{t<T_k\}}{\mathbb
E}^{{\mathbb P}} \left[\pi_{T_k}D_{T_k}\vert{\mathcal F}_t\right]. 
\label{eq:2.1}
\end{eqnarray}
Thus, at time $t$, for each cash flow that has not yet occurred we 
take its discounted risk-adjusted conditional expectation, and then 
we form the sum of such expressions to give the total value of the 
asset. 

Sometimes it is maintained that to regard share prices as being 
entirely determined by expected dividends is incorrect---that 
other factors come into play as well, such as the value implicit in 
corporate control, the value of the status of being a shareholder, 
and so on. 
In our view such ``implicit'' dividends, to the extent that they are 
relevant, and can be assigned a value, have to be modelled, 
and thus enter the valuation formula alongside the 
tangible cash flows. Sometimes it is argued that market sentiment 
is also important: indeed, it clearly is; but our view is that 
sentiment is implicit in the imperfect information the market is receiving 
concerning future cash flows; that sentiment about a future 
share price is, in essence, information concerning cash flows (both 
tangible and intangible) extending beyond the date or dates to which 
the sentiment refers. 

\begin{figure}[th]
\vspace{2cm}
\hspace{0.1cm}
\begin{picture}(12,13.5)
\unitlength8.4mm 
\put(0,0){\vector(1,0){13.5}} \put(0.9,-0.75){$\boldsymbol t$}
\put(1.0,-0.2){\line(0,1){0.4}}
\put(0,-0.75){$\bf 0$}\put(0,-0.2){\line(0,1){0.4}}
\put(2.6,-0.75){$\boldsymbol T_1$} \put(2.8,-0.2){\line(0,1){0.4}}
\put(4.8,-0.75){$\boldsymbol T_2$} \put(5,-0.2){\line(0,1){0.4}}
\put(7.3,-0.75){$\bf \ldots$} 
\put(9.8,-0.75){$\boldsymbol T_n$}\put(10,-0.2){\line(0,1){0.4}}
\put(2.2,-1.5){$\boldsymbol{\{X_{T_1}\}}$} \put(4,-1.5){$\boldsymbol{
\{X_{T_1},X_{T_2}\}}$} \put(7.3,-1.5){$\bf\ldots$}
\put(8.8,-1.5){$\boldsymbol{\{X_{T_1},\cdots,X_{T_n}\}}$}
\put(2.8,.75){\vector(0,1){1.5}} \put(2.4,2.5){$\boldsymbol{D_{T_1}}$}
\put(5,.75){\vector(0,1){1.5}} \put(4.6,2.5){$\boldsymbol{D_{T_2}}$}
\put(10,.75){\vector(0,1){1.5}} \put(9.6,2.5){$\boldsymbol{D_{T_n}}$}
\end{picture}
\vspace{1.8cm}
\caption{\small The value $S_t$ at time $t$ of a security that delivers 
the random cash flows $D_{T_1}, D_{T_2}, \cdots$, at times 
$T_1,T_2,\cdots$, is determined by the valuation formula 
(\ref{eq:2.1}). The cash flow $D_{T_1}$ is determined by a set of 
one or more independent $X$-factors $\{X_{T_1}\}$. Then 
$D_{T_2}$ is determined by 
$\{X_{T_1},X_{T_2}\}$, where $X_{T_2}$ represents a further set of 
independent $X$-factors, and so on. \label{fig:1}}
\end{figure}
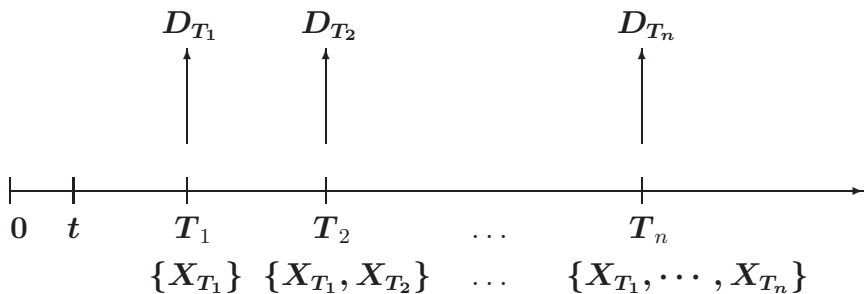

In order to apply the valuation formula we need to model the market 
filtration $\{{\mathcal F}_t\}$, as well as the pricing kernel $\{\pi_t\}$. 
In particular, it is logical to model the filtration first since the 
pricing kernel has to be adapted to the filtration. To model the 
filtration we proceed as follows. Let us introduce a set of independent 
random variables $\{X_{T_k}\}_{k=1,\ldots,n}$, which we call 
\textit{market factors} or simply ``$X$-factors''. For each $k$, the cash 
flow $D_{T_k}$ is assumed to depend on the market factors
$X^{}_{T_1},X^{}_{T_2},\cdots,X^{}_{T_k}$. Thus, in association 
with each date $T_k$
we introduce a so-called ``cash-flow function'' $\Delta_{T_k}$ such that
\begin{eqnarray}
D_{T_k}=\Delta_{T_k}(X^{}_{T_1},X^{}_{T_2},\cdots,X^{}_{T_k}). 
\label{eq:1.2}
\end{eqnarray}
For each asset, we need to model the $X$-factors,
the {\it a priori} probabilities, and the
cash-flow functions. In general, the $X$-factor associated with a
given date will be a vectorial quantity. The cash-flow diagram associated 
with a typical asset is illustrated schematically in Figure~\ref{fig:1}. 

\vspace{0.75cm}

\section{$X$-factor analysis}

Let us look at some elementary examples of cash-flow models 
based on $X$-factors. 
The first example we consider is a simple credit-risky bond, 
with two remaining coupons to be paid and no recovery on 
default. Then we have the following cash-flow structure: 
\begin{eqnarray}
D_{T_1}={c}X_{T_1}, 
\end{eqnarray}
\begin{eqnarray}
D_{T_2}=({c}+{n})X_{T_1} X_{T_2}. 
\end{eqnarray}
Here ${c}$ and $n$ denote the coupon and principal, 
respectively; and $X_{T_1}$ and $X_{T_2}$ are independent 
digital random variables
taking the values $0$ or $1$ with designated {\it a priori} 
probabilities. Evidently, if the first coupon is not paid, then
neither will the second. On the other hand, even if $X_{T_1}$ 
takes the value unity, and the first coupon is paid, the second 
coupon and the principal will not be paid unless $X_{T_2}$ 
also takes the value unity. 

The second example is a simple model for a 
credit-risky coupon 
bond with recovery. In this case the cash-flow functions are given 
as follows:
\begin{eqnarray}
D_{T_1}={c}X_{T_1}+R_{1}({c}+{n})(1-X_{T_1}),
\end{eqnarray}
\begin{eqnarray}
D_{T_2}=({c}+{n})X_{T_1}X_{T_2}+R_2({c}+{n})
X_{T_1}(1-X_{T_2}). 
\end{eqnarray}
Here $R_1$ and $R_2$ denote recovery rates. Thus if default 
occurs at the first coupon, then both the coupon and principal become 
immediately due, and a fixed fraction $R_1$ of ${c}+{n}$ is paid. 
But if default occurs at the second coupon date then the recovery rate is 
$R_2$. We observe that the $X$-factor method allows for a rather 
transparent representation of the cash-flow structure of such a 
security, and isolates the variables that underlie the various cash 
flows.

\section{Information processes}

We assume that with reference to each market factor market 
participants will have access to information, which in general is 
imperfect. 
We model the imperfect information available to market participants 
concerning a typical market factor $X^{}_{T}$ with the introduction of 
a so-called ``information process''
$\{\xi^{}_{tT}\}_{0\le t\le T}$. An information process is required to 
have the property 
\begin{eqnarray}
\xi^{}_{TT}=f(X^{}_{T}) \label{eq:3.1}
\end{eqnarray}
for some invertible function $f(x)$. This condition ensures that the 
information process ``reveals" the value of the associated market 
factor $X_T$ at time $T$. At earlier times, the value of $\xi_{tT}$ 
contains ``partial information" about the value of the $X$-factor. We 
shall come to some explicit examples of information processes shortly. 

We are now in a position to say how we model the market 
filtration. In particular, we shall assume that $\{{\mathcal F}_t\}$ is 
generated collectively by the various market information 
processes $\{\xi^{}_{tT_k}\}_{k=1,\ldots,n}$. In other words, the 
information at time $t$ is given by the following sigma-algebra:
\begin{eqnarray}
{\mathcal F}_t={\sigma}
\left[\left\{\xi^{}_{sT_k}\right\}^{}_{0\le s\le t,\ k=1,...,n}\right].
\end{eqnarray}

We thus have the following sequence of ideas: market participants are  
concerned with cash flows; cash flows are dependant 
on a set of independent market factors; market participants have 
partial access to the market factors; and this imperfect 
information generates the market filtration. 

We are left with the problem 
of taking the conditional expectation of the risk-adjusted discounted 
cash flows to generate price processes; for this purpose we 
have to model the pricing kernel. 
We assume that the pricing kernel is adapted to the market 
filtration. Thus from knowledge of the history 
of the information processes from time $0$ up to time $t$ one 
can work out the value of the pricing kernel at $t$ (see, e.g., 
\cite{[4.5]}). In a typical model the pricing kernel is given
by the discounted marginal utility of consumption of a
representative agent. It is reasonable to suppose that the 
consumption plan of the agent is adapted to the information 
filtration. The idea is that the filtration represents the flow of 
information available at each time $t$ about the relevant market 
factors, and that the consumption of the agent is determined by this 
information. In other words, the agent behaves ``rationally", always 
acting optimally on the available information, in accordance with 
appropriate criteria. There may be an idiosyncratic element to 
any given agent's consumption plan that is not adapted to the 
market filtration, and is essentially private. But the \textit{representative} 
agent has no idiosyncratic consumption.

\section{Brownian-bridge information}

For the construction of explicit models it is useful to transform to the 
risk-neutral measure ${\mathbb Q}$. This can be achieved by use of 
the pricing kernel, which we regard as specified. 
Thus for the present 
we confine the discussion to ``microeconomic'' issues: we take no 
notice of the informational notions implicit in the formulation of the 
pricing kernel, and make the 
additional simplifying assumption in what follows that the 
default-free interest-rate system is deterministic. Then the
valuation formula takes the following form:
\begin{eqnarray}\label{eq:4.1}
S_t=\sum^n_{k=1}{\mathds 1}_{\{t<T_k\}}
P_{tT_k}{\mathbb E}^{{\mathbb Q}}
\left[D_{T_k}\vert{\mathcal F}_t\right].
\end{eqnarray}
Absence of arbitrage implies that the 
discount bond system $\{P_{tT}\}_{0\le t\le T<\infty}$ is of the form
$P_{tT}=P_{0T}/P_{0t}$, 
where $\{P_{0t}\}_{0\le t<\infty}$ is the initial term structure. 

With these assumptions in place, we are in a position to 
specify a model for the information flow. For each $X$-factor $X_{T}$ 
we take the associated information process to be of the form
\begin{eqnarray}
\xi^{}_{tT}=\sigma t X^{}_{T}+\beta^{}_{tT}. \label{eq:5.3}
\end{eqnarray}
Here $\{\beta^{}_{tT}\}$ is a ${\mathbb
Q}$-Brownian bridge over the interval $[0,T]$, 
satisfying $\beta_{0T}=0$, $\beta_{TT}=0$, ${\mathbb E}
[\beta_{tT}]=0$, and
${\mathbb E}\left[\beta_{sT}\beta_{tT}\right] = s(T-t)/T$. 
The $X$-factor and the Brownian
bridge are assumed to be ${\mathbb Q}$-independent. Thus the 
Brownian bridge represents ``market noise'' and
only the ``market signal" term involving the $X$-factor contains 
true market information. The parameter $\sigma$ can be
interpreted as the ``information flow rate'' for the factor
$X^{}_{T}$. 

In the situation where we have a multiplicity of factors 
$X_{T_k}$ ($k=1,\ldots,n$), the information processes 
are taken to be of the form 
\begin{eqnarray}
\xi^{}_{tT_k}=\sigma_k t X^{}_{T_k}+\beta^{}_{tT_k},
\end{eqnarray}
where we assume that the $X$-factors and Brownian 
bridges are independent. 

The motivation for the use of a bridge to represent 
noise is intuitively as follows. We assume that initially all available 
market information is taken into account in the determination of 
prices, or, equivalently, the {\it a priori} probability laws for the market 
factors. After the passage of time, new stories circulate, 
and we model this by taking into account that the
variance of the Brownian bridge increases for the first
half of its trajectory. Eventually, the variance falls to zero 
at $T$, when the ``moment of truth'' arrives. The parameter 
$\sigma$ represents the rate at which
the true value of $X_{T}$ is ``revealed'' as time progresses.
Thus, if $\sigma$ is low, then $X_{T}$ is effectively
hidden until near the time $T$; on the other hand, if 
$\sigma$ is high, then we can think of $X_{T}$ as being
revealed quickly. If the $X$-factor is ``dimensionless'', then 
$\sigma$ has the units
\begin{eqnarray}
\sigma\sim[{\rm time}]^{-1/2}, \label{eq:c2.7}
\end{eqnarray}
and a rough measure for the timescale $\tau$ over which information 
is revealed is 
\begin{eqnarray}
\tau = \frac{1}{\sigma^2\,{\rm Var}[X_{T}]}. \label{eq:c2.8}
\end{eqnarray}
In particular, if $\tau \ll T$, then the value of $X_{T}$ will be
revealed rather early, e.g., after the passage of a few multiples 
of $\tau$. On the other hand, if $\tau\gg T$, then $X_{T}$ will 
only be revealed at the last minute, as a ``surprise''.

We remark that the information process (\ref{eq:5.3}) 
has the Markov property. This 
feature implies simplifications in the resulting models. 
In particular, on account of the relation (\ref{eq:3.1}) we find that 
the conditioning with respect to ${\mathcal F}_t$ in (\ref{eq:4.1}) can be 
replaced by conditioning with respect to the random variables 
$\xi_{tT_k}$ ($k=1,\ldots$). For a proof of the Markov property, 
see \cite{[1],[5]}.

\section{Assets paying a single dividend}

Consider an asset that pays single dividend $D_T\geq 0$ at time $T$, 
and assume that there is only one market factor $X_T$, so 
$D_T=f(X_T)$. For the moment, let us assume further that 
$f(x)=x$. Thus, we have $D_T=X_T$, where the 
market factor $X_T$ is a continuous nonnegative random variable 
with {\it a priori} ${\mathbb Q}$-density $p(x)$ for $x>0$. It follows 
by use of the Markov property of $\{\xi_{tT}\}$ that the price 
of such an asset can be written in the form
\begin{eqnarray}
S_t&=&P_{tT}{\mathbb E}\left[D_T|\xi_{tT}\right]
\nonumber \\
&=&P_{tT}\int_0^{\infty}x\,p_t(x)\,\rd x,
\end{eqnarray}
where $p_t(x)$ is the conditional density of 
$X_T$. Making use of the the Bayes formula 
one can show that $p_t(x)$ is given more explicitly 
by 
\begin{eqnarray}
p_t(x)= \frac{p(x)\exp\left[ \frac{T}{T-t}(\sigma
x\xi_{tT}-\tfrac{1}{2}\sigma^2 x^2 t)\right]}{\int^{\infty}_0
p(x)\exp\left[\frac{T}{T-t}(\sigma x\xi_{tT}-\tfrac{1}{2}\sigma^2
x^2 t)\right]\rd x}. \label{eq:1.15}
\end{eqnarray}
Thus at each time $t<T$ the price 
of the asset is determined by the random value of the information 
$\xi_{tT}$ available at that time, and is given by 
\begin{eqnarray}
S_t = P_{tT}\, \frac{\int_0^\infty x p(x)\exp\left[ \frac{T}{T-t}(\sigma
x\xi_{tT}-\tfrac{1}{2}\sigma^2 x^2 t)\right]\rd x}{\int^{\infty}_0
p(x)\exp\left[\frac{T}{T-t}(\sigma x\xi_{tT}-\tfrac{1}{2}\sigma^2
x^2 t)\right]\rd x} \label{eq:5.33}
\end{eqnarray}
The dynamics of the price process can then be obtained by an 
application of Ito's lemma, with the following result: 
\begin{eqnarray}
\rd S_t=r_t S_t\rd t+P_{tT}\frac{\sigma T}{T-t}\, 
\textrm{Var}_t\left[X_T\right]\rd W_t. \label{eq:6.4}
\end{eqnarray}
Here 
\begin{eqnarray}
\textrm{Var}_t\left[X_T\right] = \int_0^\infty x^2 p_t(x)\rd x - 
\left( \int_0^\infty x p_t(x)\rd x \right)^2
\end{eqnarray}
denotes the conditional variance of $X_T$, which by (\ref{eq:1.15})
is evidently 
given as a function of $t$ and $\xi_{tT}$. The 
$\{{\mathcal F}_t\}$-adapted process $\{W_t\}$ driving 
the dynamics of the asset in (\ref{eq:6.4}) above 
is not given exogenously, but rather is
defined in terms of the information process itself for $t<T$ by the
following formula:
\begin{eqnarray}
W_t=\xi_{tT}-\int^t_0\frac{1}{T-s}\left(\sigma
T{\mathbb E}_s\left[X_T\right]-\xi_{sT}\right)\rd s.
\end{eqnarray}
Indeed, one can verify, by use of the L\'evy criterion, that the process
$\{W_t\}$, as thus defined, is an $\{{\mathcal F}_t\}$-Brownian motion. 
Hence we see that in an information-based 
approach we can \textit{derive} the Brownian motions that drive the 
markets: they are not ``inputs'' to the model, but rather can be seen 
as arising as a ``consequence'' of the model.

\section{Geometric Brownian motion model}

A simple application of the $X$-factor technique arises in the
case of geometric Brownian motion models.
We consider a limited-liability company that makes a single cash
distribution $S_T$ at time $T$. Alternatively, think of a portfolio 
containing a single stock which will be sold off at time $T$ for 
$S_T$, with the proceeds of the sale going to the investor. We 
assume that $S_T$ has a log-normal distribution under ${\mathbb
Q}$, and can be written in the form
\begin{eqnarray}
S_T=S_0 \exp\left(rT+\nu\sqrt{T}X_T -\half \nu^2T\right),
\label{eq:zz1}
\end{eqnarray}
where the market factor $X_T$ is normally distributed with mean zero
and variance one, and where $r>0$ and $\nu>0$ are constants.
The information process $\{\xi_t\}$ is taken to be of the form 
(\ref{eq:5.3}), where in the present example 
the information flow rate is given by
\begin{eqnarray}
\sigma = \frac{1}{\sqrt{T}}.  \label{eq:zz3}
\end{eqnarray}
By use of the Bayes formula we find that the conditional probability
density is Gaussian,
\begin{eqnarray}
p_t(x) = \sqrt{\frac{T}{2\pi(T-t)}}\,\exp\left( -\frac{1}{2(T-t)}
\left( \sqrt{T}x-\xi_{tT}\right)^2\right), \label{eq:zz4}
\end{eqnarray}
and has the following dynamics:
\begin{eqnarray}
\rd p_t(x) = \frac{1}{T-t}\, \left(\sqrt{T}x-\xi_{tT}\right) p_t(x)
\rd \xi_{tT}.  \label{eq:zz5}
\end{eqnarray}
A short calculation then shows that the value of the asset in this 
example is given at time $t<T$ by
\begin{eqnarray}
S_t &=& \re^{-r(T-t)}{\mathbb E}_t[S_T] \nonumber \\ &=&
\re^{-r(T-t)} \int_{-\infty}^\infty S_0
\re^{rT+\nu\sqrt{T}x-\frac{1}{2}\,\nu^2T} p_t(x) \rd x \nonumber \\
&=& S_0 \exp\left(rt+\nu \xi_{tT}-\half\nu^2t \right) . \label{eq:zz6}
\end{eqnarray}

The surprising fact is that $\{\xi_{tT}\}$ itself turns
out to be an $\{{\mathcal F}_t\}$-Brownian motion.
Hence, writing $W_t=\xi_{tT}$ for $0\leq t\leq T$ we obtain 
the standard geometric Brownian motion model:
\begin{eqnarray}
S_t=S_0 \exp\left(rt+\nu W_t-\half\nu^2t \right).  \label{eq:zz7}
\end{eqnarray}
We see that starting with an information-based argument 
we are able to recover the familiar asset price
dynamics given by (\ref{eq:zz7}).
An important point to note is that the Brownian bridge process
$\{\beta_{tT}\}$ arises naturally in this context.
In fact, if we start with (\ref{eq:zz7}) then we can make use of the
following well-known orthogonal decomposition:
\begin{eqnarray}
W_t = \frac{t}{T}\,W_T + \left( W_t-\frac{t}{T}\,W_T\right).
\label{eq:46}
\end{eqnarray}
The second term on the right, which is independent of the first term on the
right, is a standard representation for a Brownian bridge process:
\begin{eqnarray}
\beta_{tT} = W_t - \frac{t}{T}\,W_T.
\end{eqnarray}
Then by setting $X_T= W_T/ \sqrt{T}$ and $\sigma=1/\sqrt{T}$ we find
that the right side of (\ref{eq:46}) is indeed the market
information. In other words, when it is formulated in an 
information-based framework, the
standard Black-Scholes-Merton theory can be expressed in terms of a
normally distributed $X$-factor and an independent Brownian-bridge
noise process.

\section{Pricing contingent claims}

The information-based price (\ref{eq:5.33}) of a single-dividend 
paying asset at first glance appears to be given by a rather complicated 
expression, 
suggesting perhaps that it would be impractical for use as a model 
for the pricing and hedging of contingent 
claims. However, there is a remarkable simplification involving 
a change of measure that allows one both to price and to hedge vanilla 
options. This can be seen as follows. Let us consider a 
European-style call option on 
the asset, with option maturity $t$ and strike $K$. The 
value of the option at time $0$ is given by 
\begin{eqnarray}
C_0=P_{0t}{\mathbb E}^{\mathbb Q}\left[(S_{t}-K)^+\right]. 
\label{eq:7.1}
\end{eqnarray}
Let us define a process $\{\Phi_t\}$ by the expression appearing 
in the denominator of (\ref{eq:5.33}), so 
\begin{eqnarray}
\Phi_t = \int^{\infty}_0
p(x)\exp\left[\frac{T}{T-t}(\sigma x\xi_{tT}-\tfrac{1}{2}\sigma^2
x^2 t)\right]\rd x. \label{eq:7.2}
\end{eqnarray}
Then it can be shown that $\{\Phi_t^{-1}\}$ is a positive 
${\mathbb Q}$-martingale, which can be used to change the 
probability measure from ${\mathbb Q}$ to a new measure 
${\mathbb B}$. Under the measure ${\mathbb B}$, which we 
call the ``bridge measure'', the information process itself is a  
Brownian bridge. More precisely, under ${\mathbb B}$ the process 
$\{\xi_{sT}\}_{0\leq s\leq t}$ has the law of a Brownian 
bridge spanning the interval $[0,T]$, restricted to $[0,t]$. That is to 
say, $\{\xi_{sT}\}_{0\leq s\leq t}$ is ${\mathbb B}$-Gaussian with 
mean zero and covariance ${\rm cov}[\xi_{aT},\xi_{bT}]=a(T-b)/T$ 
for $0\leq a\leq b\leq t$. The initial value of the option is thus given by 
\begin{eqnarray}
C_0&=&P_{0t}{\mathbb E}^{\mathbb B}\left[
\left( P_{tT} \int^{\infty}_0\!\! x p(x)
\re^{\frac{T}{T-t}(\sigma x\xi_{tT}-\frac{1}{2}\sigma^2 x^2 t)}\rd x 
\right. \right. \nonumber \\ && \qquad \qquad \left. \left. 
-K \int^{\infty}_0\!\! p(x)
\re^{\frac{T}{T-t}(\sigma x\xi_{tT}-\frac{1}{2}\sigma^2 x^2 t)}\rd x
\right)^+\right]. \label{eq:7.3}
\end{eqnarray}

It can be shown that the asset price is a monotonically increasing 
function of the value of $\xi_{tT}$. It follows that there is a unique 
critical level $\xi^*$ for the information 
such that the expression inside the max-function in (\ref{eq:7.3}) 
is positive. It 
follows that the option price can be written in terms of a single 
integration involving the normal distribution function: 
\begin{eqnarray}
C_0=P_{0t} \int_0^\infty p(x) \left( P_{tT}x - K \right) N\left( 
\frac{\xi^*-\sigma x t}{\sqrt{t(T-t)/T}} \right) 
\rd x .
\label{eq:7.3x}
\end{eqnarray}

As another example we consider the following. 
Suppose that the single cash flow $D_T$ is a 
binary random variable taking the values $\{d_0,d_1\}$ with 
\textit{a priori} probabilities $\{p_0,p_1\}$. The asset in this 
case can be thought of as a simple credit-risky discount 
bond that pays $d_1$ if there is no default, and $d_0$ if there is 
a default. A short calculation allows one to verify that 
\begin{eqnarray}
C_0=P_{0t}\Big[ p_1(P_{tT}d_1-K)N(u^+)-p_0 (K-P_{tT}d_0) N(u^-)
\Big], \label{eq:7.4}
\end{eqnarray}
where $u^+$ and $u^-$ are defined by
\begin{eqnarray}
u^{\pm}=\frac{\ln\left[\frac{p_1(P_{tT}d_1-K)} {p_0(K-P_{tT}d_0)}
\right]\pm\tfrac{1}{2}\sigma^2 (d_1-d_0)^2\tau}{\sigma\sqrt{\tau}
(d_1-d_0)}, \label{eq:7.5}
\end{eqnarray}
with $\tau=tT/(T-1)$. It can be shown that the option delta at time 
$0$, defined as usual by
\begin{eqnarray}
\delta_0=\frac{\partial C_0}{\partial S_{0}},
\end{eqnarray}
can be calculated explicitly, with the following result: 
\begin{eqnarray}
\delta_0 = \frac{(P_{tT}d_{1}-K)N(u^{+}) + (K-P_{tT}d_{0}) N(u^{-})}
{P_{tT}(d_1-d_0)}.
\end{eqnarray}
We see, therefore, that the apparent complexity of (\ref{eq:5.33}) 
does not lead to any intractability when it comes to derivatives 
pricing and hedging.

\section{Volatility and correlation}

In the case of an asset that pays multiple dividends, the price 
is determined by the conditional expectation given in equation (\ref{eq:4.1}). 
In terms of the cash-flow functions defined by (\ref{eq:1.2}) 
we thus obtain the following for the dynamics of the asset price: 
\begin{eqnarray}
\rd S_t &=& r_t\,S_t\,\rd t+\sum_{k=1}^n \Delta_{T_k}\rd 
{\mathds 1}_{\{t<T_k\}}\nonumber \\ 
&& + \sum^n_{k=1} {\mathds 1}_{\{t<T_k\}}\,P_{tT_k}\sum^k_{j=1}
\frac{\sigma^{}_j T_j}{T_j-t}\,{\rm Cov}_t\left[\Delta_{T_k}, X^{}_{T_j} 
\right]\rd W^{j}_t. \label{eq:8.1}
\end{eqnarray}
The leading term in the drift is the short rate, as one 
might expect, and there is also a term representing the downward 
jump in the asset that occurs when a dividend is paid. 
The independent $\{{\mathcal F}_t\}$-adapted Brownian 
motions $\{W^{j}_t\}$ driving the price dynamics 
are given in terms of the corresponding information processes by
\begin{eqnarray}
W^{j}_t=\xi^{}_{tT_j}-\int^t_0\frac{1}{T_j-s}\left(\sigma^{}_{j}\,T_j\,
{\mathbb E}_s\left[X_{T_j}\right]-\xi^{}_{sT_j}\right)\rd s.
\end{eqnarray}
We see that if an asset delivers one or more cash
flows depending on two or more market factors, then it will exhibit
``unhedgeable'' stochastic volatility \cite{[2],[4]}. That is to say, one would 
not expect to be able to hedge a position in an option by use of 
a position in the underlying. In general, if the asset cash flows depend 
on $n$ $X$-factors in total, then to hedge a generic derivative based 
on the given asset one will need the underlying together with $n-1$ 
options as hedging instruments, i.e. $n$ hedging instruments 
in total. One can read from (\ref{eq:8.1}) the generic form of the 
stochastic volatility implied for a given configuration of $X$-factors 
and cash-flow functions. 

It follows likewise from (\ref{eq:8.1}) that two or 
more assets will exhibit dynamic 
correlation when they share one or more $X$-factors in common. As 
a specific example of dynamic correlation let us 
consider a pair of credit-risky discount bonds. The
first bond is defined by a cash flow $D_{T_1}$ at $T_1$. The second 
is defined by a cash flow $D_{T_2}$ at $T_2>T_1$.
The cash flow structure is taken to be:
\begin{eqnarray}
D_{T_1}={n}_1 X_{T_1}+R_1 {n}_1(1-X_{T_1}), 
\end{eqnarray}
and
\begin{eqnarray}
D_{T_2}&=&{n}_2 X_{T_1}X_{T_2}+R^a_{2}
{n}_2(1-X_{T_1})X_{T_2} \nonumber \\
&& + R^b_{2}{n}_2 X_{T_1}(1-X_{T_2})+R^c_2
{n}_2(1-X_{T_1})(1-X_{T_2}).
\end{eqnarray}
Here, ${n}_1$ and ${n}_2$ denote the bond principals, and 
$X_{T_1}$ and $X_{T_2}$ are independent digital random variables. The 
possible recovery rates in the case of default are denoted 
by $R_1$, $R^a_2$, $R^b_2$, and $R^c_2$. One can have in 
mind the following story. Consider a  
factory with debt $S_t^{1}$. Across the street is a little 
restaurant with debt $S_t^{2}$. If the factory goes bust 
($X_{T_1}=0$) then so will the restaurant, because this is where 
the workers have their lunch. On the other hand, even if the 
factory is successful ($X_{T_1}=1$), the restaurant may still 
go bust on account of bad management ($X_{T_2}=0$). The 
recovery rates on the restaurant bond depend on the details of 
what goes wrong: $R^a_2$ (restaurant fails because factory 
fails); $R^b_2$ (restaurant fails on account of bad 
management); $R^c_2$ (factory fails, and bad restaurant 
management). One might expect $R^b_2>R^a_2$, since as long 
as the factory continues, the restaurant facilities could be sold 
at a good price. The worst scenario is that of $R^c_2$. 
For the dynamics of 
the first bond (the ``factory''), for which the price is 
\begin{eqnarray}
S^{1}_t=P_{tT_1}{\mathbb E}_t\left[D_{T_1}\right], 
\qquad (t<T_1),
\end{eqnarray}
we have:
\begin{eqnarray}
\rd S^{1}_t=r_t S^{1}_t\rd t+P_{tT_1}\frac{\sigma_1
T_1}{T_1-t}\,\alpha\,\textrm{Var}_t\left[X_{T_1}\right]\rd
W^{1}_t,
\end{eqnarray}
where $\alpha={n}_1(1-R_1)$. For the dynamics of the second
bond (the ``restaurant''), for which the price is 
\begin{eqnarray}
S^{2}_t=P_{tT_2}{\mathbb E}_t\left[D_{T_2}\right], 
\qquad (t<T_2),
\end{eqnarray}
we have:
\begin{eqnarray}
\rd S^{2}_t &=& r_t S^{2}_t\rd t + P_{tT_2}\frac{\sigma_1T_1}
{T_1-t}\left(\beta+\delta{\mathbb E}_t\left[X_{T_2}\right]\right)\textrm{Var}_t 
\left[X_{T_1}\right]\rd W^{1}_t \nonumber \\
&& + P_{tT_2}\frac{\sigma_2 T_2}{T_2-t}\left(\gamma+\delta
{\mathbb E}_t\left[X_{T_1}\right]\right)\textrm{Var}_t\left[X_{T_2}\right]\rd
W^{2}_t,
\end{eqnarray}
where the constants $\beta$, $\gamma$, and $\delta$ are given by 
$\beta={n}_2(R^b_2-R^c_2)$, $\gamma={n}_2(R^a_2-R^c_2)$, 
and $\delta={n}_2(1-R^a_2-R^b_2+R^c_2)$. The filtration 
$\{{\mathcal F}_s\}$ is generated by the information processes 
$\{\xi_{sT_1}\}$ and $\{\xi_{sT_2}\}$ 
associated with $X_{T_1}$ and $X_{T_2}$. The dynamics 
of the bond prices depend on a common Brownian driver 
$\{W^{1}_t\}$. The fact that the asset payoffs share a common 
$X$-factor thus gives rise to a dynamic correlation between the movements 
of the price processes $\{S^{1}_t\}$ and $\{S^{2}_t\}$. 
The instantaneous correlation between the price movements of the 
factory bond and the restaurant bond is given by the following 
expression:  
\begin{eqnarray}
\rho_t=\frac{\rd S^{1}_t\,\rd S^{2}_t}{\sqrt{\big(\rd
S^{1}_t\big)^2\big(\rd S^{2}_t\big)^2}}.
\end{eqnarray}
Hence, using the formulae for the dynamics of the two 
assets, we obtain:
\begin{eqnarray}
\rho_t=\frac{1}{\sqrt{\psi_t}}
\frac{\sigma_1T_1}{T_1-t}\left(\beta+\delta
{\mathbb E}_t\left[X_{T_1}\right]\right)
{\rm Var}_t\left[X_{T_1}\right] , 
\label{eq:7.10}
\end{eqnarray}
where 
\begin{eqnarray}
\psi_t &=& \left(\frac{\sigma_1
T_1}{T_1-t}\right)^2\left(\beta+\delta\,{\mathbb E}_t\left[
X_{T_2}\right]\right)^2\left({\rm Var}_t\left[X_{T_1}\right]\right)^2
\nonumber \\ && 
+\left(\frac{\sigma_2 T_2}{T_2-t}\right)^2\left(\gamma+
\delta\,{\mathbb E}_t \left[X_{T_1}\right]\right)^2\left(
{\rm Var}_t\left[X_{T_2}\right]\right)^2. 
\end{eqnarray}
We see from (\ref{eq:7.10}) 
that we are able to calculate explicitly the dynamics of the
correlation between the movements of the two asset prices.

\section{Amount of information about the future cash flow
contained in the price process} 

Since we are modelling the flow of information 
in an explicit manner, we are able to quantify how much information 
regarding the value of the cash flow $D_T$ is contained in the value 
$\xi_t$ at time $t$ of the associated 
information process. For simplicity we shall in the discussion that 
follows assume that the cash flow $D_T$ takes the discrete values 
$\{d_i\}_{i=1,\ldots,n}$ with \textit{a priori} probabilities $\{p_i\}_{i=1,
\ldots,n}$. A reasonable measure for quantifying the information 
content is given by the mutual information $J(\xi_t ,D_T)$ between 
the two random variables, which in the present context is given by 
the expression
\begin{eqnarray}
J(\xi_t,D_T) = \sum_{i=1}^n \int_{-\infty}^\infty \rho(\xi,i)\, 
\ln\left( \frac{\rho(\xi,i)}{\rho(\xi)\rho(i)}\right) \rd \xi, \label{eq:8.1x}
\end{eqnarray}
where
\begin{eqnarray}
\rho(\xi,i) = \frac{\rd}{\rd \xi}\, {\mathbb Q}\Big[
(\xi_t<\xi) \cap (D_T=d_i)\Big]
\end{eqnarray}
is the joint density of the random variables $\xi_t$ and $D_T)$,
and $\rho(\xi)$ and $\rho(i)$ are the respective marginal probabilities. 
By use of the relation
\begin{eqnarray}
{\mathbb Q}\Big[ (\xi_t<\xi) \cap (D_T=d_i)\Big] = {\mathbb Q}(\xi_t<\xi
| D_T=d_i)\,{\mathbb Q}(D_T=d_i)
\end{eqnarray}
we deduce that
\begin{eqnarray}
\rho(\xi,i) = p_i \frac{1}{\sqrt{2\pi t(T-t)/T}}\, \exp
\left(-\half \frac{(x-\sigma d_it)^2}{t(T-t)/T}\right),
\label{eq:8.4}
\end{eqnarray}
since conditional on $D_T=d_i$ the random variable $\xi_t$ is
normally distributed with mean $\sigma t d_i$ and variance
$t(T-t)/T$. From (\ref{eq:8.4}) the marginal densities
\begin{eqnarray}
\rho(\xi) = \sum_{i=1}^n \rho(\xi,i) \quad {\rm and} \quad 
\rho(i) = \int_{-\infty}^\infty \rho(\xi,i) \rd \xi
\end{eqnarray}
can be deduced. In particular, $\rho(i)=p_i$. 
By substituting (\ref{eq:8.4}) in (\ref{eq:8.1x}), the 
information about the cash flow $D_T$ contained in $\xi_t$ can 
be determined.

From an information-theoretic point of view, two processes
related through an invertible function, thus sharing 
the \textit{same} filtration, in general possess \textit{different}
information content. On the other hand, since what is 
observed in the market is the price $S_{t}$, which is an 
invertible function of $\xi_t$, it is more relevant to determine the 
mutual information $J(S_{t},D_T)$, that is, the amount of information
about the future cash flow contained in the market price. It can be 
shown that in the present context we have 
$J(S_{t},D_T)=J(\xi_{t},D_T)$.

\section{Information disparity and statistical arbitrage}

So far we have assumed that market participants have equal
access to information, but one can ask what happens if some traders
are more ``informed" than others. Suppose we consider a
financial product that pays a single cash flow $D_T$ at 
time $T$. We can think of this product as a kind of bond. The general
market trader has access to an information process concerning $D_T$,
but there are also ``informed" traders who have access 
to one or more additional information processes concerning $D_T$. The 
informed trader is thus in some sense able to make 
a ``better estimate" of the value of the asset. 

\begin{figure}[t]
\begin{center}
  \includegraphics[scale=0.65]{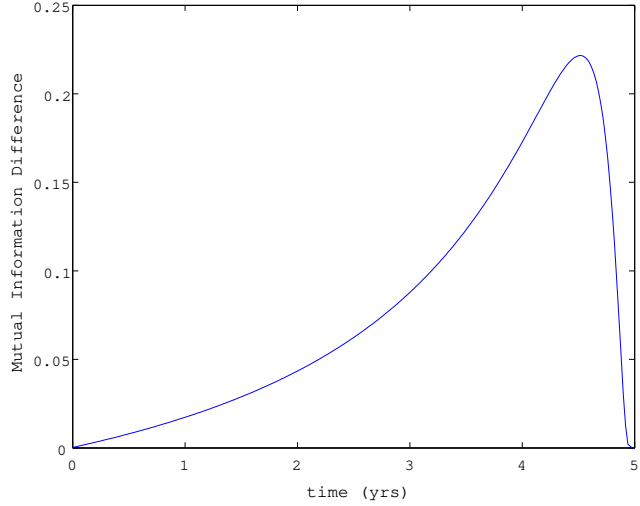}
  \caption{\small Mutual information difference. The additional information
held by an informed trader over that of the market is nonnegative.
The parameters are set to be $d_1=0$, $d_2=1$, $p_1=0.2$, $p_2=0.8$,
$T=5$, $\sigma=0.25$, $\sigma'=0.45$, and $\rho=0.15$ (the correlation 
between $\beta_{tT}$ and $\beta'_{tT}$).
  \label{fig:2}
  }
\end{center}
\end{figure}

To be more specific, let us suppose that 
while the general market trader has 
access to the information $\xi_t=\sigma t D_T + \beta_{tT}$, an 
informed trader has access in addition, say, to the information 
$\xi'_t=\sigma' t D_T + \beta'_{tT}$, where 
$\{\beta_{tT}^\prime\}$ may or may not be correlated with the market 
noise $\{\beta_{tT}\}$. Thus,  the information source for the informed 
trader is given by $\{{\mathcal G}_t\}= {\sigma}( 
\{\xi_s,\xi'_s\}_{0\leq s\leq t})$. The use 
of such extra information can, but need not, represent ``insider 
trading'' in the usual sense. 
That is, it may be that the informed trader merely has better 
access to (and better computer power for the purpose of digesting) 
the vast domain of 
publicly available information. Since we have introduced an 
information measure regarding impending cash flows, we can 
quantify the excess information held by the informed 
trader above that held by general market traders. This is 
measured by 
the difference of the mutual information $\Delta J$. In 
figure~\ref{fig:2} we plot an example of $\Delta J$, indicating the 
way in which the excess information held by the informed trader 
changes in time. 

Given that the informed trader is on average ``more knowledgeable"
than the general market trader, it is natural to ask how the
informed trader can take advantage of the situation to seek so-called 
``statistical arbitrage" opportunities. We assume that the
informed trader operates on a relatively small scale, and that the
actions of this trader do not significantly influence the
market. Suppose we consider a trading strategy such
that at some designated time $t\in(0,T)$ a market trader purchases a
bond if, and only if, at that time the bond price $S_{t}$ is
greater than the quantity $KP_{tT}$ for some specified constant $K$. 
Once a bond is purchased, it is held to maturity. The
informed trader follows the same rule, but makes a better estimate
of the value of the bond, and hence purchases the bond iff ${\tilde
S}_{t}>KP_{tT}$, where 
\begin{eqnarray}
{\tilde S}_{t}=P_{tT}{\mathbb E}[D_T|{\mathcal G}_t]. 
\end{eqnarray}
The significance of ${\tilde S}_t$ is that it 
represents the price that the informed trader knows that the 
market as a whole would make if the market as a whole had the 
same knowledge as the informed trader. 

\begin{figure}[t]
\begin{center}
  \includegraphics[scale=0.65]{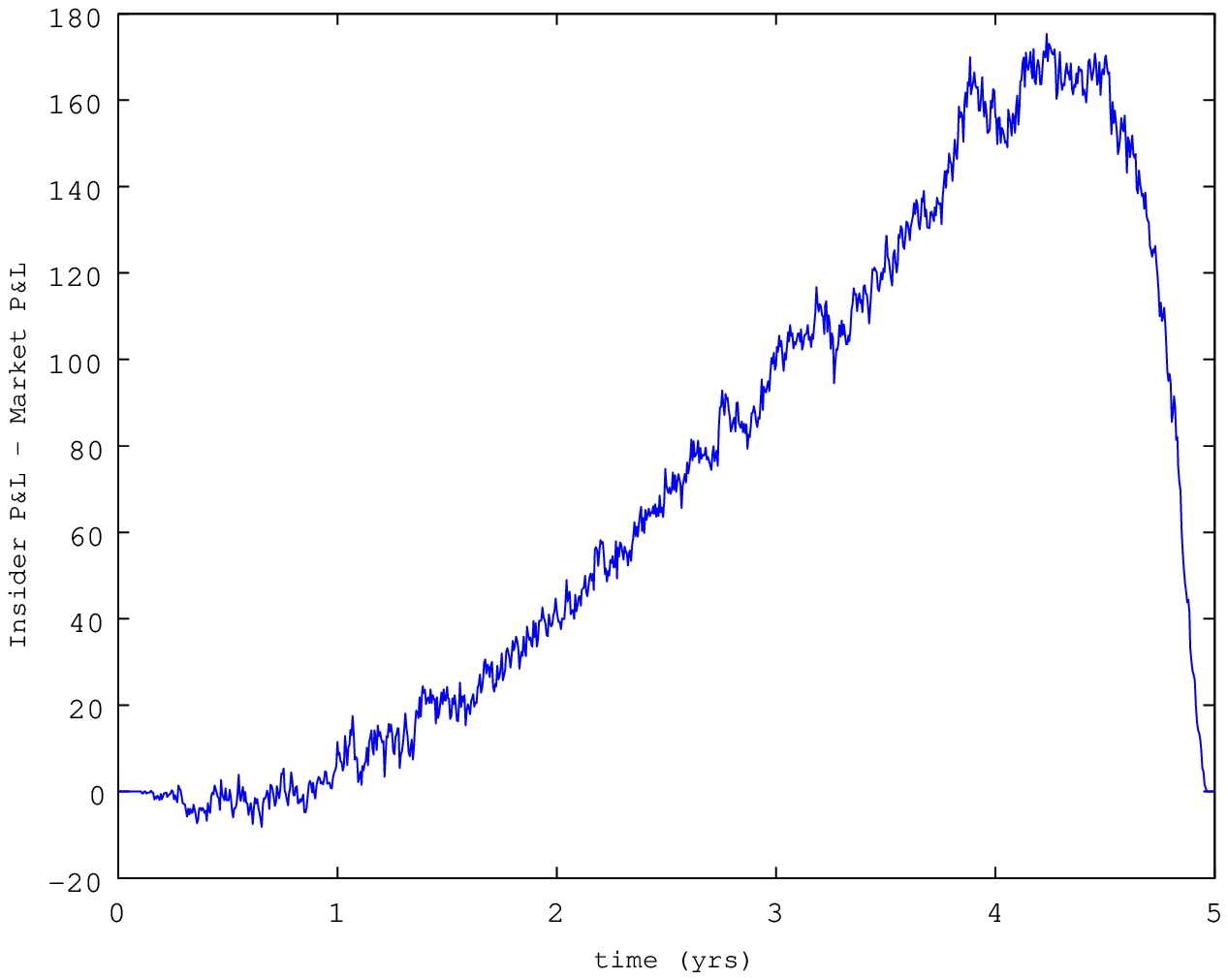}
  \caption{\small The P\&L difference for digital bonds. At each time 
traders purchase the bond if and only if the valuation of the bond
is greater than a specified threshold. The general market trader
buys if $S_{t}>KP_{tT}$, whereas the informed trader uses the
criterion ${\tilde S}_{t}>KP_{tT}$. The difference in profit and
loss between the informed trader and the general market trader is
plotted, based on $2000$ realisations, when the \textit{a priori}
probability of default is $p_1=0.2$. Other parameters are set to be
$d_1=0$, $d_2=1$, $T=5$, $\sigma=0.25$, $\sigma'=0.45$, 
$\rho=0.15$, and $K=0.7$.
  \label{fig:3}
  }
\end{center}
\end{figure}

That such a strategy leads to a statistical arbitrage opportunity
for the informed trader can be seen as follows \cite{[7]}. 
We assume that the initial position of a 
trader is zero, i.e. purchase of a bond at $t$ requires
borrowing the amount $S_{t}$ at that time, and repaying the amount
$P_{tT}^{-1}S_{t}$ at $T$. Thus the value of a general market trader's
portfolio at $T$ is
\begin{eqnarray}
V_T={\mathds 1}{\{S_{t}>KP_{tT} \}} (D_T-P_{tT}^{-1}S_{t}),
\end{eqnarray}
whereas the value of the informed trader's portfolio at $T$ is
\begin{eqnarray}
{\tilde V}_T={\mathds 1}{\{{\tilde S}_{t}>KP_{tT}\}}
(D_T-P_{tT}^{-1}S_{tT}).
\end{eqnarray}
Consider the present value $P_{0T} {\mathbb E}[\Delta V_T]$ of a
security that delivers a cash flow equal to the excess profit or
loss
\begin{eqnarray}
\Delta V_T={\tilde V}_T- V_T
\end{eqnarray}
generated by the strategy of the informed trader. By the
tower property we have
\begin{eqnarray}
{\mathbb E}[\Delta V_T]={\mathbb E}[{\mathbb E}[\Delta V_T|{\mathcal
G}_t]].
\end{eqnarray}
However, 
\begin{eqnarray}
{\mathbb E}[\Delta V_T|{\mathcal G}_t] = P_{tT}^{-1}\left({\mathds 1}
{\{{\tilde S}_{t}>KP_{tT}\}} - {\mathds 1}{\{S_{t}>KP_{tT}\}} \right)
\left({\tilde S}_{t}-S_{t} \right), \label{eq:7.4}
\end{eqnarray}
since the random variables $S_{t}$ and ${\tilde S}_{t}$ are 
${\mathcal G}_t$-measurable. If ${\tilde S}_{t}>S_{t}$ then
\begin{eqnarray}
{\mathds 1}{\{{\tilde S}_{t}>KP_{tT}\}} - {\mathds 1}{\{S_{t}>KP_{tT}\}} 
\geq 0; 
\end{eqnarray}
whereas if ${\tilde S}_{t}<S_{t}$ then
\begin{eqnarray}
{\mathds 1}{\{{\tilde S}_{t}>KP_{tT}\}} - {\mathds 1}{\{S_{t}>KP_{tT}\}} 
\leq 0.
\end{eqnarray}
It follows that ${\mathbb E}[\Delta V_T|{\mathcal G}_t]>0$ with 
probability greater than zero, and therefore ${\mathbb E}
[\Delta V_T]>0$. We know that according to the 
usual no-arbitrage arguments the present value
of the payoff of the strategy of the general market trader must be
zero. It follows that the informed trader can execute a transaction at
zero cost that has positive value: this is what we mean by
``statistical arbitrage". A simulation study of the profit arising 
from this trading strategy is shown in figure~\ref{fig:3}, indicating 
a close correspondence with the excess information 
held by the informed trader shown in figure~\ref{fig:2}.

\section{Price formation in inhomogeneous markets}

The idea of ``informed trading'' can be extended to a market 
that has a number of traders operating in it, all more or less on 
an equal footing, but where different traders have access to 
different information. In other words, there is an inhomogeneous 
information flow in the market. This line of thinking leads naturally 
to the consideration of price formation in such a market.

Let us consider, as an example, a market with two traders, labelled 
``Trader 1'' and ``Trader 2''. As before, there is a single asset, with a single 
dividend $D_T$ paid at time $T$. The traders have access to 
separate sources 
of information about $D_T$, given respectively by $\xi^1_t = 
\sigma_1t D_T+ \beta^1_{tT}$ and $\xi^2_t = \sigma_2 t D_T + 
\beta^2_{tT}$. 
Here the Brownian bridges $\{\beta_{tT}^1\}$ and 
$\{\beta^2_{tT}\}$ are assumed, for 
simplicity, to be independent. Trader 1 works 
out the price
\begin{eqnarray}
S_t^1 = P_{tT} {\mathbb E}[D_T|\xi^1_t]
\end{eqnarray}
that he knows the market would have made had the market 
possessed the information generated by $\{\xi^1_t\}$. Likewise, 
Trader 2 works out the price
\begin{eqnarray}
S_t^2 = P_{tT} {\mathbb E}[D_T|\xi^2_t]
\end{eqnarray}
that she knows the market would have made had the market 
possessed the information generated by $\{\xi^2_t\}$. 

Traders 1 and 2 are unaware of each other's prices, but can 
gain information by trading. The trading process works as 
follows. Each trader makes a spread about their price. Letting 
$ 0< \phi^- < 1 < \phi^+$, we set
\begin{eqnarray}
S_t^{1\pm} = \phi^\pm S_t^1
\end{eqnarray}
for the buy price $S_t^{1-}$ and sell price $S_t^{1+}$ made by 
Trader 1 at time $t$. Thus
Trader 1 is willing to buy at a price slightly below his
information-based valuation $S_t^1$, and is willing to sell at a price
that is slightly above that valuation. Likewise Trader 2 is willing to
buy at a price slightly below her information-based valuation $S_t^2$,
and is willing to sell at a price that is slightly above that valuation: 
\begin{eqnarray}
S_t^{2\pm} = \phi^\pm S_t^2.
\end{eqnarray}
We assume that there is an exchange that
continuously monitors the prices made by the traders. 
The exchange effects a trade of some fixed size
when the buy price of one of the traders reaches the level of the
sell price of the other trader. That is to say, a trade takes place
when
\begin{eqnarray}
S_t^{1-} = S_t^{2+} \quad \textrm{or} \quad S_t^{1+} = S_t^{2-}.
\end{eqnarray}
When a trade occurs, at that moment each trader learns the
price of the other, and as a consequence can back out the value 
of the corresponding information process. Therefore, when a 
trade occurs, the traders each briefly have access to both pieces 
of information, and are thus in a position to make a better price, 
namely that given by
\begin{eqnarray}
S_t^{1,2} = P_{tT} {\mathbb E}[D_T|\xi_t^1,\xi^2_t].
\end{eqnarray}
We conclude that immediately after a trade the information-based
prices made by each of the traders will jump to the same level, and
that the {\it a priori} probability distribution for $D_T$ will be updated 
correspondingly.

Once the trade is concluded, the link between the two traders is
lost, and each trader again has access only  to their own
information source. Starting from the same price, the prices made by
the two traders diverge as they receive different information going
forward. A further trade will then occur when the buy price of one
of the traders next hits the sell price of the other trader. 

At the time the trade is executed, the joint information can be 
embodied in the value of an effective information process 
$\{{\hat\xi}_t\}_{0\le t\le T}$ given by 
\begin{eqnarray}
{\hat\xi}_t=\sqrt{(\sigma_{1})^2+(\sigma_{2})^2} \, t D_T +
\frac{\sigma_1\, \beta^{1}_{tT}+\sigma_2\,\beta^{2}_{tT}}
{\sqrt{(\sigma_{1})^2 +(\sigma_{2})^2}}.
\end{eqnarray}
We note that $\{{\hat\xi}_t\}$ is indeed an information process, 
since it can be written in the form 
\begin{eqnarray}
{\hat\xi}_t=\hat{\sigma} t D_T + \hat{\beta}_{tT},
\end{eqnarray}
where
\begin{eqnarray}
\hat{\sigma}=\sqrt{(\sigma_{1})^2 +(\sigma_{2})^2}, \quad 
{\rm and} \quad \hat{\beta}_{tT}=
\frac{\sigma_1\,\beta^{1}_{tT}+\sigma_2\,
\beta^{2}_{tT}}{\sqrt{(\sigma_{1})^2 +(\sigma_{2})^2}}.
\end{eqnarray}
One can show that $\{\hat{\beta}_{tT}\}$ is a Brownian bridge and 
is independent of 
$D_T$. Thus, immediately after the trade is executed, the price
$S^{1,2}_{t}$ made by both traders is of the form
\begin{eqnarray}
S^{1,2}_{t}=P_{tT}\,\frac{\int^{\infty}_0 x\,
p(x)\exp\left[\frac{T}{T-t}\left(\hat{\sigma}\,x\,
{\hat\xi}_t-\tfrac{1}{2}\,\hat{\sigma}^2\,
x^2\,t\right)\right]\rd x}{\int^{\infty}_0
p(x)\exp\left[\frac{T}{T-t}\left(\hat{\sigma}\,x\,
{\hat\xi}_t-\tfrac{1}{2}\,\hat{\sigma}^2
\,x^2\,t\right)\right]\rd x}.
\end{eqnarray}

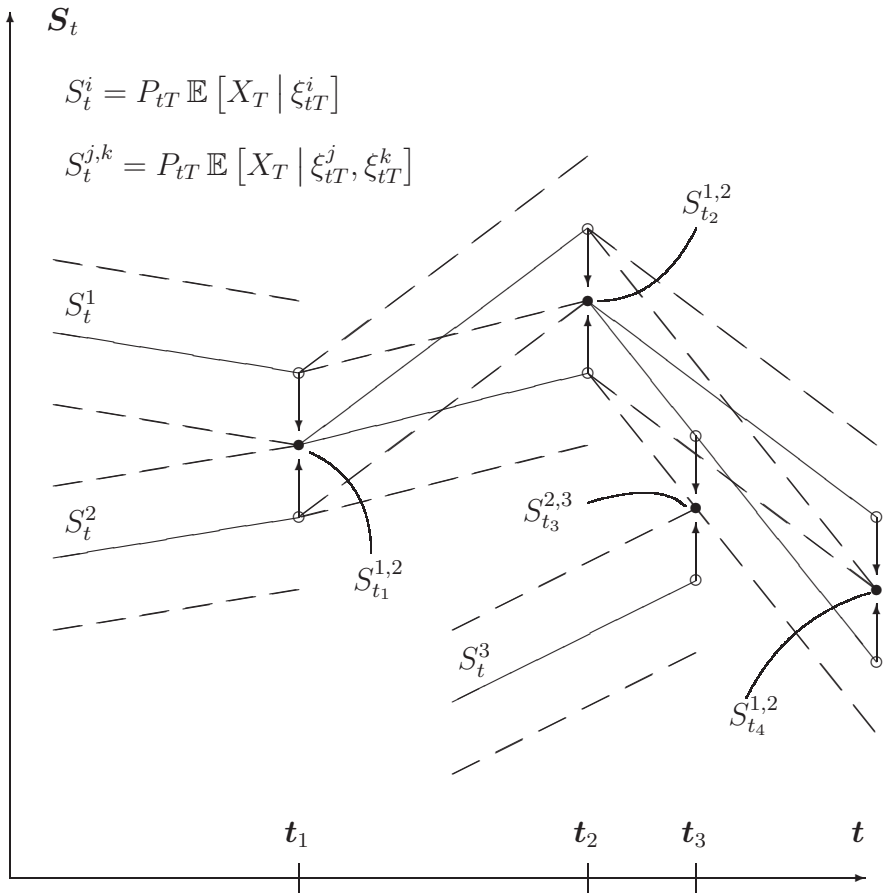
\begin{figure}[th]
\vspace{12cm} 
\hspace{0.1cm}
\begin{picture}(12,13.5)
\unitlength9.6mm 
\put(0,0){\vector(1,0){11.85}}\put(11.65,0.5){$\boldsymbol t$}
\put(0,0){\vector(0,1){12}} \put(0.5,11.75){$\boldsymbol S_t$}
\put(3.8,0.5){$\boldsymbol t_1$} \put(4,-0.2){\line(0,1){0.4}}
\put(7.8,0.5){$\boldsymbol t_2$} \put(8,-0.2){\line(0,1){0.4}}
\put(9.3,0.5){$\boldsymbol t_3$} \put(9.5,-0.2){\line(0,1){0.4}}
\put(.75,10.75){$S_{t}^{i}=P_{tT}\,{\mathbb E} 
\left[X_T\,\big\vert\,\xi^{i}_{tT} \right]$}
\put(.75,9.75){$S_{t}^{j,k}=P_{tT}\,{\mathbb E}
\left[X_T\,\big\vert\,\xi^{j}_{tT},\xi^{k}_{tT}
\right]$}
\put(.75,7.8){$S^{1}_t$}
\multiput(4.0,8.0)(-0.6,0.1){6}{\line(-6,1){0.4}}
\put(4,7){\line(-6,1){3.4}}\put(4,7){\circle{0.15}}
\multiput(4.0,6.0)(-0.6,0.1){6}{\line(-6,1){0.4}}
\put(4,7){\vector(0,-1){0.8}}
\put(4,6){\circle*{0.15}}
\qbezier(4.15,5.9)(5,5.5)(5,4.5)\put(4.75,4){$S^{1,2}_{t_1}$}
\put(0.75,4.8){$S^{2}_t$}
\multiput(4.0,6.0)(-0.6,-0.1){6}{\line(-6,-1){0.4}}
\put(4,5){\line(-6,-1){3.4}}\put(4,5){\circle{0.15}}
\multiput(4.0,4.0)(-0.6,-0.1){6}{\line(-6,-1){0.4}}
\put(4,5){\vector(0,1){0.8}}

\multiput(8.0,8.0)(-0.6,-0.15){7}{\line(-4,-1){0.4}}
\put(8,7){\line(-4,-1){4}}\put(8,7){\circle{0.15}}
\multiput(8.0,6.0)(-0.6,-0.15){7}{\line(-4,-1){0.4}}
\put(8,7){\vector(0,1){0.8}}
\put(8,8){\circle*{0.15}}
\qbezier(8.15,8)(9,8)(9.5,9)
\put(9.3,9.22){$S^{1,2}_{t_2}$}
\multiput(8.0,8.0)(-0.6,-0.45){7}{\line(-4,-3){0.4}}
\put(8,9){\line(-4,-3){4}}\put(8,9){\circle{0.15}}
\multiput(8.0,10.0)(-0.6,-0.45){7}{\line(-4,-3){0.4}}
\put(8,9){\vector(0,-1){0.8}}
\multiput(8.0,9.0)(0.6,-0.45){7}{\line(4,-3){0.4}}
\put(8,8){\line(4,-3){4}}\put(12,5){\circle{0.15}}
\multiput(8.0,7.0)(0.6,-0.45){7}{\line(4,-3){0.4}}
\put(12,5){\vector(0,-1){0.8}}
\put(12,4){\circle*{0.15}}
\qbezier(11.9,3.95)(10.75,3.6)(10.2,2.5)
\put(9.95,2.12){$S^{1,2}_{t_4}$}
\multiput(8.0,9.0)(0.46,-0.575){9}{\line(4,-5){0.3}}
\put(8,8){\line(4,-5){4}}\put(12,3){\circle{0.15}}
\multiput(8.0,7.0)(0.46,-0.575){9}{\line(4,-5){0.3}}
\put(12,3){\vector(0,1){0.8}}

\put(6.2,2.9){$S^{3}_t$}
\multiput(9.5,5.125)(-0.6,-0.3){6}{\line(-2,-1){0.375}}
\put(9.5,4.125){\line(-2,-1){3.375}}\put(9.5,4.125){\circle{0.15}}
\multiput(9.5,3.125)(-0.6,-0.3){6}{\line(-2,-1){0.375}}
\put(9.5,4.125){\vector(0,1){0.8}}
\put(9.5,5.125){\circle*{0.15}}
\put(9.5,6.125){\circle{0.15}}\put(9.5,6.125){\vector(0,-1){0.8}}
\qbezier(9.35,5.15)(9,5.5)(8,5.2)
\put(7.1,4.95){$S^{2,3}_{t_3}$}
\end{picture}
\caption{\small Schematic illustration of information-based trading. An 
exchange executes a trade when the sell price of Trader $i$ 
matches the buy price of Trader $j$ (dashed lines meet 
at filled dots). At the execution time $t_n$, both traders have access 
to each other's valuations $S^{i}_{t_n}$ (empty
circles) and thus to each other's information. As a
consequence, the traders are able to update their valuations and 
obtain the common price $S^{j,k}_{t_n}$. After
the trade has been executed and the respective asset valuations have
been updated, the traders ``separate", and return to their 
individual valuations. The respective valuations may drift in different
directions. The traders will get in contact again as soon as the exchange
notifies them that the respective sell and buy prices have matched
again. Such an information-based trading mechanism can be extended 
to multiple traders, as suggested in the illustration. \label{fig:4}}
\end{figure}

As an example, let us consider the case of a digital 
payout taking the values 0 and 1 with {\it a priori} 
probabilities $p_0$ and $p_1$. We consider the case in 
which the information flow rates are the same, so we set 
$\sigma_1=\sigma_2=\sigma$. Then for the valuations we 
have
\begin{eqnarray}
S_t^{1}=\frac{p_1\,\exp\left[\frac{T}{T-t}\left(\sigma\xi^{1}_t-
\tfrac{1}{2}\,\sigma^2 t\right)\right]}{p_0+p_1\,\exp\left[
\frac{T}{T-t}\left(\sigma\xi^{1}_t-\tfrac{1}{2}\,\sigma^2
t\right)\right]}
\end{eqnarray}
and
\begin{eqnarray}
S_t^{2}=\frac{p_1\,\exp\left[\frac{T}{T-t}\left(\sigma\xi^{2}_t-
\tfrac{1}{2}\,\sigma^2t\right)\right]}{p_0+p_1\,\exp\left[
\frac{T}{T-t}\left(\sigma\xi^{2}_t-\tfrac{1}{2}\,\sigma^2
t\right)\right]}.
\end{eqnarray}
For the spreads, we assume $\phi^+=1+\delta$ and $\phi^-=1-\delta$
where $\delta$ is small. If Trader 1 is the buyer then the condition
for a trade is
\begin{eqnarray}
(1-\delta)S^{1}_t=(1+\delta)S^{2}_t.
\end{eqnarray}
Given his knowledge $\xi^{1}_t$, Trader 1 can use the
condition to work out the value of $\xi_t^{2}$. In particular,
suppose that
\begin{eqnarray}
\xi^{2}_t=\xi^{1}_t+\varepsilon_t,
\end{eqnarray}
where $\varepsilon_t$ is small. Then a calculation shows that
$\varepsilon_t$ is given, to first order, by
\begin{eqnarray}
\varepsilon_t=-\frac{2\,\delta(T-t)}{\sigma\,T\left(1-
{\mathbb E}\big[X_T\,\vert\,\xi^{1}_t\big]\right)}.
\end{eqnarray}

The general situation, where there are a number of traders present
in the market, and where the asset cash flows depend on a number of
market factors, is very rich. It is evident that in the broad
picture there is no universal filtration, nor a universal pricing
measure. Nevertheless, by exchanging information through trading
activity, market participants can maintain a ``law of reasonable
price range'' if not a ``law of one price''. 

Certainly, the notion that there is 
a universal market filtration is unrealistic. What counts is not merely 
``access in principle'' to information, but rather ``access in practice''. 
Perhaps some broader version of market efficiency will survive, taking 
into account the cost of such access (cf. \cite{G}).
A subscription to the Wall Street Journal is not free, nor is a Bloomberg 
terminal. Access to vast information providers such as Google and 
Yahoo may seem free or nearly so, but from a broader perspective 
this is not so---someone pays, in cash or kind. What is the market 
price of information? And how does this depend on the ``information 
about the information''? For the answers to these questions we must 
await the development of new models.

\vspace{0.25cm} 

\noindent {\bf Acknowledgements}. The authors are grateful to seminar
participants at the Fourth General Conference on Advanced Mathematical
Methods in Finance, Alesund, Norway, May 2009, the Workshop on 
Incomplete Information in Mathematical Finance, Chemnitz, Germany, 
June 2009, the Graduate School of Management, Kyoto, October 2009, 
and the Kyoto Institute of Economic Research, January 2010, for helpful 
comments. Part of this work was carried out while LPH was visiting the 
Aspen Center for Physics (September 2009), and Kyoto University (August, 
October 2009). DCB and LPH thank HSBC, Lloyds TSB, Shell 
International, and Yahoo Japan for research support.


\begin{thebibliography}{99.}

\bibitem{[1]} D.~C.~Brody, L.~P.~Hughston \& A.~Macrina (2007) 
``Beyond hazard rates: a new framework for credit-risk modelling'' 
In {\em Advances in Mathematical Finance, Festschrift volume in 
honour of Dilip Madan}. R.~Elliott, M.~Fu, R.~Jarrow and Ju-Yi~Yen, 
eds. (Basel: Birkh\"auser).

\bibitem{[2]} D.~C.~Brody, L.~P.~Hughston \& A.~Macrina (2008)
``Information-based asset pricing'' \textit{Int. J. Theor. Appl.
Fin.} \textbf{11}, 107-142.

\bibitem{[3]}  D.~C.~Brody, L.~P.~Hughston \& A.~Macrina (2008) 
``Dam rain and cumulative gain'' \textit{Proc. Roy. Soc. Lond. } 
A\textbf{464} 1801-1822.

\bibitem{[7]} D.~C.~Brody, M.~H.~A.~Davis, R.~L.~Friedman \& 
L.~P.~Hughston (2009) ``Informed traders'' \textit{Proc. Roy.  Soc. 
Lond.} A\textbf{465}, 1103-1122.

\bibitem{G} S.~J.~Grossman \& J.~E.~Stiglitz (1980) 
``On the impossibility of informationally efficient markets'' {\em Am. 
Econ. Rev.} \textbf{70} 393-408. 

\bibitem{[6]} L.~P.~Hughston \& A. Macrina (2008) ``Information, Inflation,
and Interest" In  {\em Advances in Mathematics of Finance}, Banach
Center Publications \textbf{83} (Institute of Mathematics, Polish
Academy of Sciences).

\bibitem{[4]} A.~Macrina (2006) {\em An Information-Based 
Framework for Asset Pricing: X-factor theory and its Applications}. 
PhD thesis, King's College London. 

\bibitem{[4.5]} A.~Macrina \& P.~A.~Parbhoo (2010) ``Securities pricing 
with information-sensitive discounting'' In {\em Recent Advances in Financial 
Engineering 2009}, Proceedings of the KIER-TMU International Workshop on 
Financial Engineering 2009 (Singapore: World Scientific Publishing).

\bibitem{[5]} M.~Rutkowski \& N.~Yu (2007) ``On the 
Brody-Hughston-Macrina approach to modelling of defaultable 
term structure"  \textit{Int. J. Theor. Appl. Fin.} \textbf{10}, 557-589.

\end{thebibliography}
\end{document}